# Observation of Quantum-Criticality-Class Crossover at the $LaAlO_3/KTaO_3$ (111) Interface

Jia Liu,[1,*] Long Cheng,[1,*†] Mingyue Zhang,[1] Junkun Zha,[1,2] Fei Ye,[1] Xiaofang Zhai[1,†]

[1] *School of Physical Science and Technology, Shanghai Tech University, Pudong, Shanghai 201210, China.*

[2] *Department of Physics, University of Science and Technology of China, Hefei, Anhui 230026, China*

* *These authors contributed equally to this work.*

† *Corresponding authors: chenglong1@shanghaitech.edu.cn, zhaixf@shanghaitech.edu.cn.*

**Abstract**

In two-dimensional (2D) limit, the quantum fluctuation is significantly enhanced which could induce a quantum phase transition. Investigating the quantum criticality is an effective approach to elucidate the underlying physics of 2D superconductivity. Here we report the observation of different universality classes of quantum criticality at the superconducting $LaAlO_3/KTaO_3$ (111) interface, i.e. the normal quantum Griffiths singularity (QGS) and the anomalous QGS. The switch of the quantum criticality class is observed from the normal QGS in a lower $T_c$ sample with weaker spin-orbit coupling (SOC) to the anomalous QGS in a higher $T_c$ sample with stronger SOC. Moreover, owing to the remarkable SOC strength, the higher $T_c$ sample exhibits an unprecedentedly enhanced quantum fluctuation at an unusually elevated temperature approaching the Berezinskii-Kosterlitz-Thouless (BKT) transition temperature $T_{BKT}$. This work provides comprehensive recognition of the different quantum criticality classes in the same superconducting $LaAlO_3/KTaO_3$ (111) system, which deepens the understanding of the superconducting mechanism of interfacial superconductors.

## Introduction

The 2D superconductivity has recently emerged as the new frontier of discovering exotic physical phenomena, such as the Majorana fermion [1] , the quantum bosonic metal [2,3], the Higgs mode [4,5], etc. In 2D limit, the pronounced quantum fluctuation can induce quantum phase transitions and tailor the phase boundary by mutual interplay with thermal fluctuations, disorders, etc. In realistic superconducting materials, quenched disorder is inevitable, which could enhance the quantum fluctuation by formation of superconducting rare regions [6]. In 2D superconductors, quenched disorder leads to the divergent dynamical critical exponent towards zero temperature, which has been theoretically predicted as QGS [7]. Recently, QGS has been experimentally confirmed in many 2D disordered superconductors, where the onset critical magnetic field increases with decreasing temperature [6,8-13]. Moreover, the anomalous QGS with onset critical fields decreasing toward zero temperature has also been observed [14].

In transitional metal oxides (TMOs), competition and coupling of multiple degrees of freedom constitute the background of many exotic phenomena. For example, in high $T_c$ cuprates, competition between magnetism and superconducting is considered key to the mechanism behind the unconventional superconductivity [15]. Investigating the pair-forming/breaking characters near the quantum phase transition has become an important path to achieve insights on the superconducting mechanism in high $T_c$ cuprates [16,17] and other superconducting systems [18,19]. The superconducting two-dimensional electron gas (2DEG) at oxide interfaces, represented by the well-known $LaAlO_3/SrTiO_3$ system [20,21], is a distinct family of superconductors belonging to both TMO and 2D superconductor communities. Recently, the $KTaO_3$-based 2DEG [22,23] has been found to have almost one order-of-magnitude higher $T_c$ (~2 K) than the $LaAlO_3/SrTiO_3$ (~300 mK). In both systems, defects such as oxygen vacancies and metal substitutions not only serve as the carrier donors but also act as the quenched disorders [22], which obscures the superconducting mechanism. Furthermore, the inversion symmetry breaking at the interface promises the giant SOC, which induces mixed spin singlet and triplet pairing channels [24,25]. So far, the mechanism responsible for the $KTaO_3$ interfacial superconductivity is still under intensive study [22,23,25-28]. It is tempting to investigate the quantum phase transition in order to find out roles of different degrees of freedom, such as SOC, in manipulating the $KTaO_3$-based superconductivity.

Here we systematically study the quantum criticality of the superconducting 2DEG at the $LaAlO_3/KTaO_3$ (111) interface. The $LaAlO_3/KTaO_3$ samples with different $T_c$ were fabricated by pulsed laser deposition with varied growth conditions. Based on the thorough low temperature transport measurements, we reveal that both the $T_c$ and SOC increase with the strength of the disorder scattering. More strikingly, we observe a switch between two different universality classes of the superconducting quantum criticality. Specifically, the lower $T_c$ sample with weak SOC shows a normal-like QGS behavior, featuring the enhanced critical magnetic field near zero temperature. On the contrary, the higher $T_c$ sample with strong SOC exhibits a rarely-seen anomalous QGS, which presents a maximal critical field near the $T_{BKT}$ and an outward-buckled phase boundary. Our observations provide strong evidence that the intertwined strong SOC and high degree of disorder scattering are intimately connected to the enhanced superconductivity with extremely strong quantum fluctuation.

## Results

The $LaAlO_3/KTaO_3$ heterostructures were fabricated by growing $LaAlO_3$

overlayers on $KTaO_3$ (111) substrates using pulsed laser deposition. Samples with systematically tuned 2DEG properties were achieved by varying only the growth temperature (Methods and Fig. S1-3 in Supplementary Materials [29]). Fig. 1(a) shows the Hall bar geometry adopted for electrical transport measurements, with comprehensive fabrication details described in Methods and Fig. S4 [29]. The temperature dependent resistances under zero field of two representative samples are shown in Fig. 1(b). The $T_c$ is defined at the half resistance of the normal state is 2.08 K and 1.43 K for the two samples respectively. The $T_{BKT}$ is also estimated with the power law $V \propto I^{\alpha}$ $(\alpha = 3)$ behavior from the temperature-dependent *I-V* curves, which is determined to be 1.88 K for $KTaO_3$-#1 and 1.27 K for $KTaO_3$-#2 in Fig. S5 [29].

The residual sheet resistance at 10 K is 3693 Ω/□ for $KTaO_3$-#1 and 2037 Ω/□ for $KTaO_3$-#2. The larger residual sheet resistance implies the enhanced disorder scattering of carriers. Meanwhile, $KTaO_3$-#1 shows a lower residual resistance ratio (RRR) and a lower $k_F l$ value with $k_F$ being the Fermi wave vector and $l$ being the mean free path in Fig. S6 and Table S1 [29]. Both RRR and $k_F l$ values are widely used to characterize the strength of scattering, and the lower RRR and $k_F l$ values are associated with the stronger scattering of carriers. Thus, the above results indicate that the strong disorder scattering of carriers is likely related to the enhancement in superconductivity. Note these samples display linear Hall effects without observable nonlinear contributions from multi-band carriers (Fig. S2 [29]). Hence the transport parameters have been derived using the single band model, with the understanding that the multi-band character of the Ta $t_{2g}$ manifold may impart subtle corrections (see details in Table S1 [29]).

Figs. 1(c) and 1(d) show the field-dependent resistance $R(B)$ at different temperatures of the two samples. When the temperature is much below $T_{BKT}$, the superconductivity of the system is inhibited gradually with the increase of the magnetic field, which results in a 'U' shape $R(B)$ curve. Near $T_{BKT}$ the $R(B)$ curve exhibits a 'V' shape. Above $T_{BKT}$ the $R(B)$ curve has a weak dependence on the field. We observe crossing points of the magnetoresistance isotherms occurring at critical magnetic fields near 1.4 T and 0.5 T in $KTaO_3$-#1 and #2, respectively. Furthermore, we obtain the temperature-dependent resistance at different magnetic fields from Figs. 1(c), 1(d), which is shown in Figs. 1(e), 1(f). We found the two samples present different superconductor-metal transition (SMT) behaviors. The $KTaO_3$-#2 shows a normal SMT with monotonic phase boundary separating the regions of $dR/dT > 0$ and $dR/dT < 0$ below and above the critical field around 0.5 T (Fig. 1(f)). For $KTaO_3$-#1, when applying magnetic field higher than 1.3 T, the resistance firstly increases with decreasing temperature reaching a maximum around 2.5 K and then decreases at lower temperature. More intriguingly, when the applied magnetic field reaches 1.5 T, the $R(T)$ curve shows a reentrant behavior. Thus the higher $T_c$ and lower $T_c$ samples exhibit different types of quantum criticality when crossing the critical field.

To thoroughly study the quantum critical behavior, we zoom in the temperature-dependent magnetoresistance near the critical field. We find the $R(B)$ curves of $KTaO_3$-#1 cross each other in a relatively large transition region instead of a single critical point, as shown in Fig. 2(a). Such a behavior has been ascribed as the fingerprint proof of the QGS [6]. In comparison, $KTaO_3$-#2 shown in Fig. 2(c) exhibits a much narrower transition region. Figs. 2(b) and 2(d) show the temperature dependence of the crossing magnetic field $B_c$ extracted from the $R(B)$ curves at every two adjacent temperatures. It is obvious that the $B_c$ dependence on the temperature is drastically different between the two samples. $B_c$ of $KTaO_3$-#2 exhibits an increasing trend with decreasing temperature, complying with the normal QGS behavior [6,8-13,30]. Oppositely, $KTaO_3$-#1 exhibits an exceptional decreasing trend of $B_c$ with decreasing temperature below $T_c$, which is ascribed to the anomalous QGS [14]. The normal QGS occurs in $KTaO_3$-#2 (1.43 K), which is similar to $T_c$ (1.3 K) of the previously reported EuO/$KTaO_3$ sample with normal QGS [13]. Additionally, more superconducting samples have been measured and the anomalous QGS behavior was reproduced in samples with $T_c$ higher than 1.5 K (Fig. S7 and S8 [29]).

To analyze the quantum critical phenomenon with series of crossing points, we use the finite-size scaling [31,32] analysis to determine the critical parameter of the two samples. The resistance takes the scaling form $R(B,T) = R_c f\left[(B - B_c)/T^{1/z\nu}\right]$, where $R_c$ is the resistance at the crossing field $B_c$. $f$ is an arbitrary function of $B$ and $T$ with $f[0] = 1$. $z$ is the dynamical critical exponent and $\nu$ is the correlation length exponent. The scaling form is rewritten as $R(B,t)/R_c = f[(B - B_c)t]$ with $t = (T/T_0)^{-1/z\nu}$, and $t$ can be obtained by collapsing multiple $R(B,t)$ curves of a single crossing field $B_c$ within a temperature range from $T_0$ to $T$. Figs. 3(a)-(d) show representative groups of isothermal curves before and after the data collapsing of the two samples. Fig. 3(a) shows a group of data with $B_c = 1.40$ T and $R_c = 3583\ \Omega/\square$ of $KTaO_3$-#1 in the temperature range from 1.05 K to 1.2 K. After the finite-size scaling the data collapse onto a bivalue curve as shown in Fig. 3(c). The inset of Fig. 3(c) shows the logarithmic dependence of $t$ and $T$, from which the $z\nu = 12.97$ is obtained in the temperature range of 1.05-1.2 K. Figs. 3(b) and 3(d) show an example of $KTaO_3$-#2 with similar analysis. The derived dynamical critical exponent $z\nu$ is 1.42 in the temperature range of 0.85-1.1 K. More examples of the analysis in different temperature range can be found in Fig. S9 [29], and choice of the temperature range size does not affect the results. The *R(B)* measurements have been repeated using low-pass filters (Fig. S10 [29]) and on spatially-separated devices of a single sample (Fig. S11 [29]), which rule out possible artifacts due to noise or sample inhomogeneity.

The obtained $z\nu$ values in the full temperature range below $T_c$ are shown in Figs. 3(e) and 3(f) for $KTaO_3$-#1 and #2, respectively. As shown in Fig. 3(e), $z\nu$ grows rapidly as temperature decreases and diverges with the temperature approaching zero. The divergent behavior of $z\nu$ can be well described by the activated scaling law $z\nu \propto |B - B_C^*|^{-0.6}$ [6,8] for both samples. It indicates that the infinite-randomness quantum critical point emerges at $B_C^*$ when the temperature approaches zero. In agreement to the anomalous QGS, the crossing magnetic field tends to a minimum $B_C^*$ of 1.33 T. As for $KTaO_3$-#2, $z\nu$ also increases as temperature decreases and diverges with the temperature approaching zero. But the crossing magnetic field approaches a maximum $B_C^*$ of 0.52 T, which agrees to the typical QGS. The above quantitative analyses further confirm the QGS and the anomalous QGS in the lower $T_c$ and higher $T_c$ samples, respectively.

In Figs. 4(a) and 4(b), we sketch the schematic phase diagram of the higher $T_c$ and lower $T_c$ samples respectively. Each phase diagram contains regimes of superconducting with vortex lattice, superconducting fluctuation with QGS or anomalous QGS and the normal metal. Strikingly, for the higher $T_c$ sample, the strongest superconducting fluctuation appears near $T_{BKT}$ when the quantum fluctuation is competitive with the thermal fluctuation, since the onset critical field (crossing field) $B_c$ reaches the maximum (1.73 T). Below $T_{BKT}$, the quantum fluctuation overtakes the thermal fluctuation, while $B_c$ decreases with decreasing temperature and approaches 1.33 T at zero temperature. As a result, the superconducting fluctuation regime deeply protrudes into the normal metal regime. Different from the higher $T_c$ sample, the lower $T_c$ sample exhibits normal-like QGS, which shows that $B_c$ increases with the decreasing temperature, and approaches a maximum of 0.52 T at zero temperature. Nevertheless, relatively weak superconducting fluctuation with a small peak is observed slight above the $T_c$. For both samples, we derive the upper critical field $B_{c2}^{GL}$ using the Ginzburg-Landau mean-field formula [33]. It is found that $B_{c2}^{GL}$ of both samples is close to the quantum critical field $B_c^*$ when the temperature approaches zero. Overall, the phase diagrams demonstrate that the lower and higher $T_c$ samples exhibit distinctly different quantum fluctuation characters. Especially, the higher $T_c$ sample shows the unprecedentedly enhanced quantum fluctuation at a high temperature near $T_{BKT}$.

**Discussion**

Although QGS has been found in a number of 2D superconductors with quenched disorders [6,8-13], the anomalous QGS has been rarely discovered. Previously, Y. Liu *et al*. [14] reported the observation of the anomalous QGS in ultrathin Pb films, which is explained by the effect of pronounced superconducting fluctuations with strong SOC. But the anomalous outward-buckled superconducting phase boundary in the Pb system appears at the temperature far below the $T_c$, in contrary to the current work. The remarkable difference implies that the disorder scattering and SOC in the $LaAlO_3/KTaO_3$ system play more profound roles in the superconducting transition. In the close family of the $SrTiO_3$ and other $KTaO_3$ based 2DEG superconductors, only normal QGS has been found though both in principle exhibit strong SOC.

To quantitively evaluate the strength of SOC, we perform magnetoresistance measurements in the normal state at 12 K in both samples, and adopt the Maekawa-Fukuyama (MF) model to deduce the SOC parameters [26]. The results are shown in supplementary Fig. S12 and Table S1 [29]. It is found that $KTaO_3$-#1 exhibits a much stronger effective SOC field $B_{so}$ (~3.3 T) and SOC splitting energy $\Delta_{so}$ (~40.4 meV) than $KTaO_3$-#2 with $B_{so}$ ~0.9 T and $\Delta_{so}$ ~22.6 meV. Furthermore, the SOC strength has also been estimated in the superconducting state using modified Werthamer-Helfand-Hohenberg (WHH) model for thin films [34]: $B_{c2\parallel}(0\mathrm{K}) = \sqrt{\frac{1.76\hbar k_B T_c}{[3\mu_B^2 \tau_{so} + D(d_{sc}e)^2/3}}$. $\tau_{so}$ is the spin orbit relaxation time ($\Delta_{so} = \hbar/\tau_{so}$). $D$ is the diffusion constant obtained from the slope of the out of plane upper critical field: $[-d(B_{c2\perp})/dT]_{T=T_c} = 4k_B/\pi De$. The $d_{sc}$ is the superconducting thickness, which was obtained by fitting $B_{c2\perp}$ and $B_{c2\parallel}$ using the 2D Ginzburg-Landau (G-L) model (Fig. S13-14 [29]). We estimate $B_{c2\parallel}(0\mathrm{K})$ using the G-L model to be 19.4 T for $KTaO_3$-#1 and 6.4 T for $KTaO_3$-#2, which are both beyond the Pauli limit (3.87 T and 2.98 T). Consequently the $\tau_{so}$ and $\Delta_{so}$ are derived from the WHH model to be 0.012 ps and 57.3 meV for $KTaO_3$-#1, and 0.046 ps and 14.2 meV for $KTaO_3$-#2. Thus, the higher $T_c$ sample indeed exhibits significantly enhanced SOC as compared to the lower $T_c$ sample.

## Conclusion

In conclusion, the quantum phase transition of the superconducting 2DEG at the $LaAlO_3/KTaO_3$ interface is systematically studied by the low temperature transport measurements on samples with different disorder scattering strength. The results show that the more disordered sample possesses the higher $T_c$ and stronger SOC strength. More importantly, we find a crossover of quantum criticality class from the normal QGS in the lower $T_c$ sample to the anomalous QGS in the higher $T_c$ sample, inseparable from the enhancement of the SOC. Our findings not only deepen the insight into the exotic quantum criticality of the superconducting $LaAlO_3/KTaO_3$ interface, but also open a new frontier of researches in disorder-tuned or SOC-enriched quantum phase transitions in 2D superconductors.

## Acknowledgement

We acknowledge in-depth discussions with Jiandi Zhang, Yulin Chen, Rui Peng, Yi Liu, Jiawei Luo, Guangming Zhang and Fu-Chun Zhang. The work was financially supported by National Key R&D Program of China (Nos. 2022YFA1403000, 2023YFA1406301), the National Science Foundation of China (Nos. 12104305, 52072244), the Science and Technology Commission of Shanghai Municipality (Nos. 22TS1401200). The research used resources from Analytical Instrumentation Center (#SPST-AIC10112914) and Soft Matter Nanofab (SMN180827) in ShanghaiTech University.

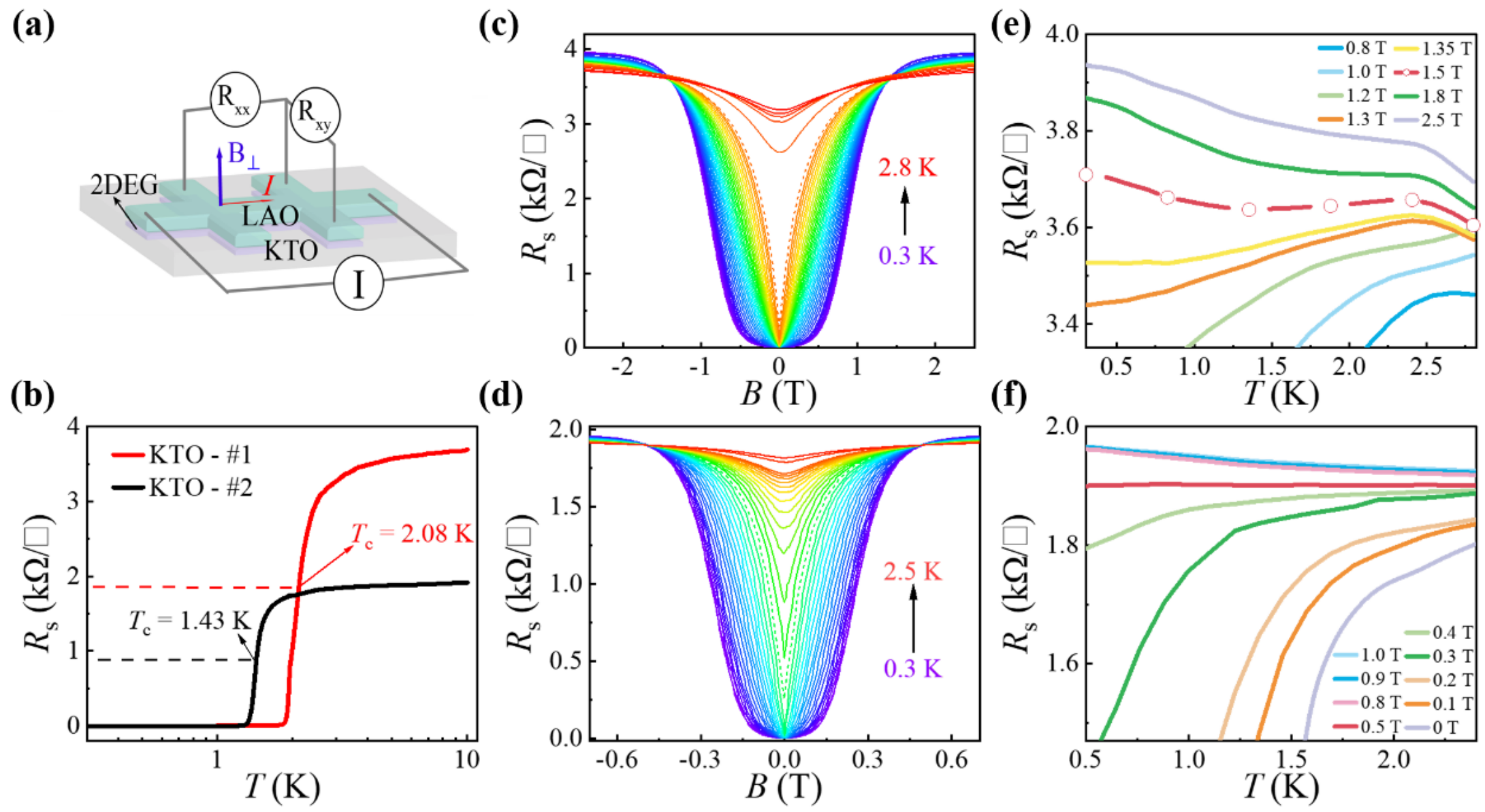

FIG.1 Transport measurements for $LaAlO_3/KTaO_3$ (111) samples. (a) Schematic of the Hall bar device and the measurement geometry. (b) Temperature-dependent sheet resistance below 10 K under zero magnetic field. $T_c$ (defined as $R(T_c) = 0.5 \times R(10\text{ K})$) is indicated by the red arrow for Sample labeled as KTO-#1 and the black arrow for Sample labeled as KTO-#2, respectively. The isotherm $R(B)$ curves measured at different temperatures of (c) KTO-#1 and (d) KTO-#2. The dashed curves indicate $R(B)$ curves near $T_{BKT}$. The isomagnetic $R$-$T$ dependence under various perpendicular magnetic fields of (e) Sample KTO-#1 and (f) Sample KTO-#2. The superconducting reentrant behavior in Sample KTO-#1 is highlighted by the red line with hollow circles.

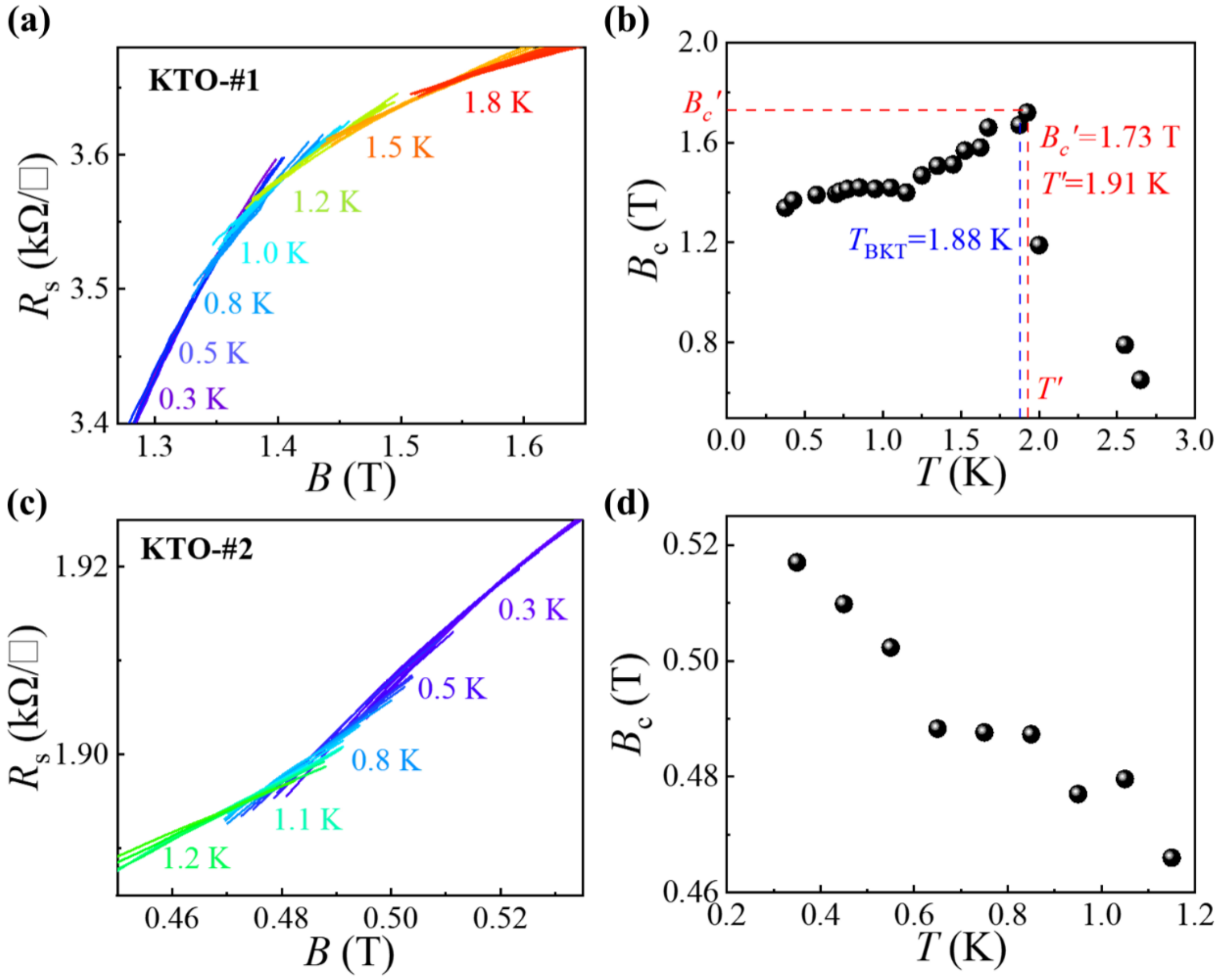

FIG.2 Anomalous QGS and normal QGS behaviors. (a) The crossing region of isotherms $R(B)$ curves of KTO-#1 from 0.3 K to 1.8 K. (b) The temperature dependence of crossing point $B_c$ of every two adjacent $R(B)$ curves deduced from (a). (c) The crossing region of isothermal $R(B)$ curves of KTO-#2 from 0.3 K to 1.2 K. (d) The temperature dependence of crossing point $B_c$ of every two adjacent $R(B)$ curves deduced from (c).

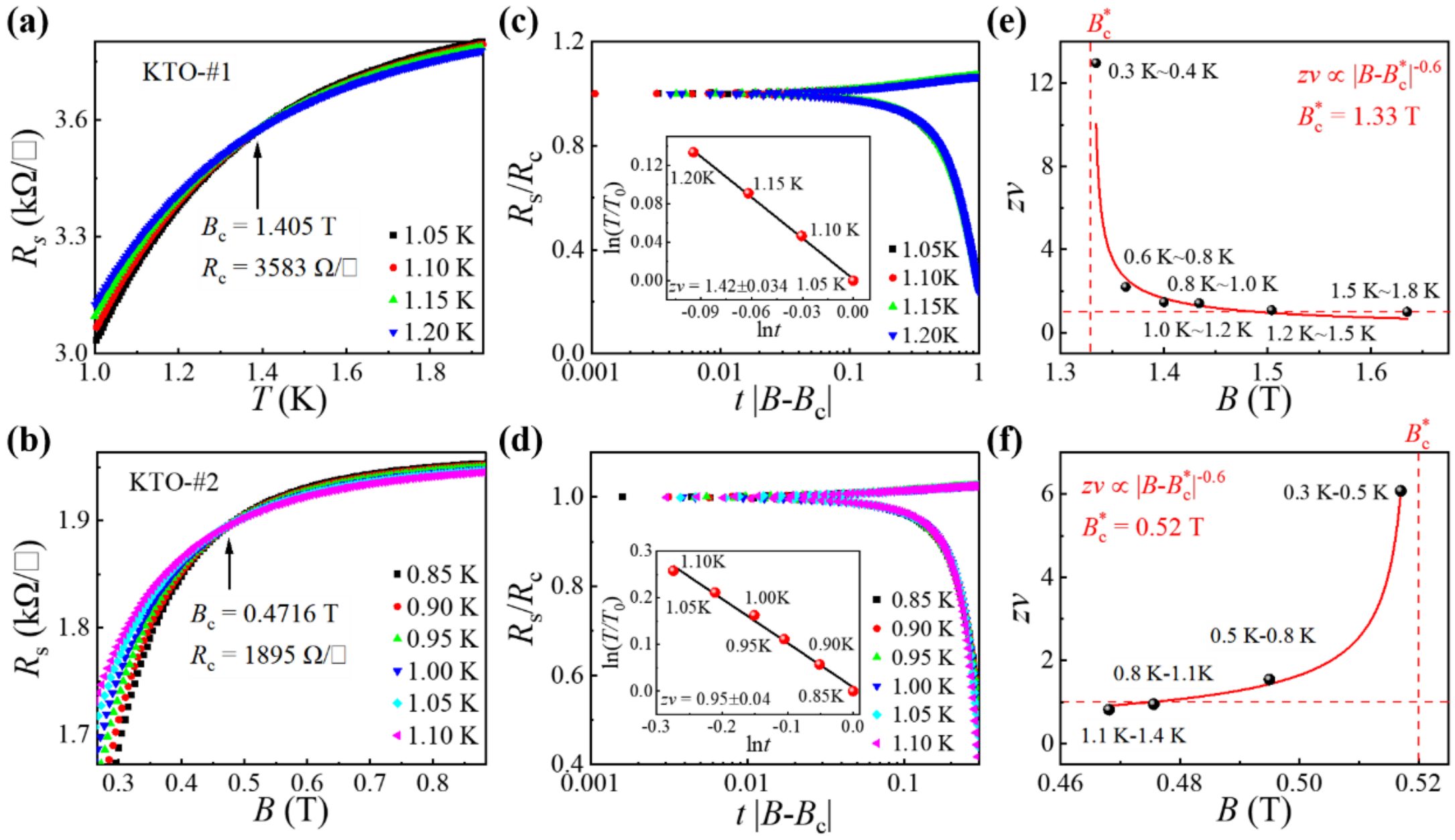

FIG.3 The Finite-size scaling analysis of representative groups for both samples with the anomalous QGS and normal QGS behaviors. (a-d) The isothermal $R(B)$ curves of (a) KTO-#1 in the temperature range of 1.05-1.2 K and (b) KTO-#2 in the temperature range of 0.85-1.1 K. (c), (d) The corresponding normalized sheet resistance of (a) and (b) as a function of $t|B-B_c|$, where $t = (T/T_0)^{-1/z\nu}$. The insets show the linear fitting between $\ln(T/T_0)$ and $\ln t$. The $z\nu$ values are obtained from slopes of the fitting lines. (e), (f) The divergent behavior of $z\nu$ with temperature approaching zero and magnetic field tending to (e) a minimal $B_c^*$ in KTO-#1 corresponding to anomalous QGS and (f) a maximal $B_c^*$ in KTO-#2 corresponding to normal QGS. $B_c^*$ is deduced from the activated scaling law $z\nu \propto |B - B_c^*|^{-0.6}$.

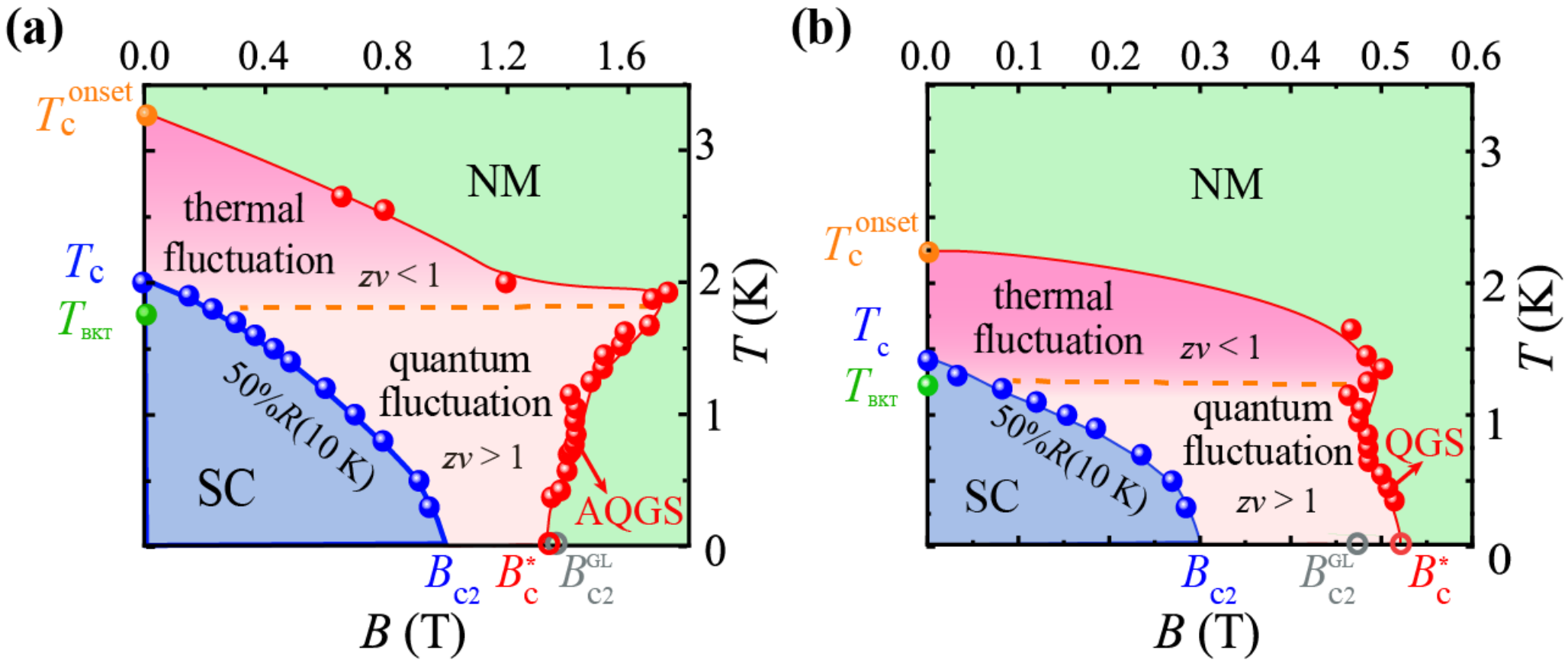

FIG.4 The Phase diagram for (a) the higher $T_c$ sample with anomalous QGS and (b) the lower $T_c$ sample with normal-like QGS, respectively. The plots are based on the transport data. Red spheres and blue spheres represent the crossing points $B_c$ of the isothermal $R(B)$ curves and the upper critical magnetic field $B_{c2}$. The orange sphere indicates the onset temperature of superconducting transition with $R(T_c^{onset}) = 0.9 \times R(10\text{ K})$. The red hollow indicates the infinite-randomness quantum critical point $B_c^*$ at zero temperature. The gray hollow indicates $B_c^{GL}$ extrapolated from the Ginzburg-Landau formula. SC and NM represent superconducting state and normal metal, respectively. When $z\nu > 1$, thermal fluctuation dominates the phase transition; when $z\nu < 1$, quantum fluctuation takes over. In both diagrams, the division line $z\nu = 1$ is found at $T_{BKT}$.